\documentclass[aps,prl,twocolumn,showpacs,superscriptaddress,groupedaddress,nofootinbib]{revtex4-1}  % for review and submission
\usepackage{graphicx}  % needed for figures
\usepackage{dcolumn}   % needed for some tables
\usepackage{bm}        % for math
\usepackage{aas_macros}
\usepackage{amssymb}   % for math
\usepackage{amsmath}
\usepackage{color}
\begin{document}
\title{The effect of positron-alkali metal atom interactions in the diffuse ISM}
	\author{Fiona H. Panther}
		%\email{fiona.panther@anu.edu.au}
		\affiliation{
			Research School of Astronomy and Astrophysics \\ Australian National University \\ Canberra 2611, Australia\\}
            \author{Ivo R. Seitenzahl}
            \affiliation{School of Physical, Environmental and Mathematical Sciences, University of New South Wales,\\ Australian Defence Force Academy, Canberra 2612, Australia}
            \author{Roland M. Crocker}
            \affiliation{Research School of Astronomy and Astrophysics \\ Australian National University \\ Canberra 2611, Australia\\}
            \author{Joshua R. Machacek}
\affiliation{Plasma Research Laboratory, Research School of Physics and Engineering, Australian National University, Acton, Canberra, 2601}
\author{Dan J. Murtagh}
\affiliation{Plasma Research Laboratory, Research School of Physics and Engineering, Australian National University, Acton, Canberra, 2601}
\affiliation{Stefan Meyer Institute for Subatomic Physics, Boltzmanngasse 3,1090 Vienna, Austria}
            \author{Thomas Siegert, Roland Diehl}
            \affiliation{Max-Planck-Institut f\"{u}r extraterrestrische Physik, Giessenbachstr. 1, D-85741 Garching, Germany}
            
\bibliographystyle{apsrev}
\date{\today}

	\begin{abstract}
		In the Milky Way galaxy, positrons, which are responsible for the diffuse $511\,\mathrm{keV}$ gamma ray emission observed by space-based gamma ray observatories, are thought to annihilate predominantly through charge exchange interactions with neutral hydrogen. These charge exchange interactions can only take place if positrons have energies greater than $6.8\,\mathrm{eV}$, the minimum energy required to liberate the electron bound to the hydrogen atom and then form positronium, a short-lived bound state composed of a positron-electron pair. Here we demonstrate the importance of positron interactions with neutral alkali metals in the warm interstellar medium (ISM). Positrons may undergo charge exchange with these atoms at any energy. In particular, we show that including positron interactions with sodium at solar abundance in the warm ISM can significantly reduce the annihilation timescale of positrons with energies below $6.8\,\mathrm{eV}$ by at least an order of magnitude. We show that including these interactions in our understanding of positron annihilation in the Milky Way rules out the idea that the number of positrons in the Galactic ISM could be maintained in steady state by injection events occurring at a typical periodicity $>\mathrm{Myr}$.
	\end{abstract}
	\maketitle

	\section{\label{sec:intro}Introduction}
    In 1931, Paul Dirac predicted the existence of a particle with the same mass, but opposite charge, to the electron \citep{Dirac31}. The existence of the anti-electron, or positron, was confirmed by Carl Anderson a year later \citep{Anderson32}. While the annihilation of positrons through direct interaction with electrons was predicted by Dirac \citep{Dirac31}, positrons may also form a short-lived bound state composed of a positron and electron, known as positronium, before subsequently annihilating. The existence of positronium was predicted in 1934 \cite{Mohorovicic1934}, but was not confirmed experimentally until the 1950's \citep{Deutsch51}. Since the 1930's, positrons have been observed in a variety of astrophysical environments. The presence of positrons is inferred through the detection of characteristic gamma ray emission at $m_e c^2 \sim 0.5\,\mathrm{MeV}$ from the co-annihilation of electrons and positrons. We now see positron annihilation in the diffuse interstellar medium (ISM) \cite{Knoedelseder05,Weidenspointner08,Siegert16}, solar flares \citep{Murphy05}, associated with the synthesis of $\beta^+$ unstable radionuclides in supernovae \citep{Churazov14}, and in microquasar flares \cite{Siegertmicroquasars}. Futhermore, the production of positrons in astrophysical environments can also be inferred through the observed presence of the parent nuclei of $\beta^+$ unstable radionuclides such as $^{26}$Al, $^{22}$Na, $^{44}$Ti and $^{56}$Ni, whose decay lines are observed through both X-ray and gamma ray measurements (see \citep{Diehl11} for an overview).\\
    
In the early 1970's, emission of gamma rays at $\sim0.5\,\mathrm{MeV}$ from the central regions of the Milky Way was first detected by balloon-borne spectrometers \citep{Johnson72,Leventhal1978} and subsequently confirmed by a number of space-based missions \citep{Purcell97,Knoedelseder05,Weidenspointner08,Siegert16}. Recent observations of diffuse annihilation gamma rays across the Milky Way suggest an annihilation rate of  $\sim5\times10^{43}$ positrons per second \citep{Siegert16}. These positrons are constrained to be injected into the ISM at mildly relativistic energies ($<3-7\,\mathrm{MeV}$)\citep{Aharonian81,Beacom06} due to the absence of a gamma ray continuum at energies $>511\,\mathrm{keV}$ produced when relativistic positrons annihilate in flight with free electrons. Positron lifetimes are split into two phases: the ``in-flight" phase, when the kinetic energy of the positrons is greater than $w_\mathrm{therm} = 3kT_e/2$ (where $T_e$ is the electron temperature) and the ``thermalized" phase, when the positron kinetic energy drops below $w_\mathrm{therm}$ \citep{Guessoum05}.\\
In the ISM, positrons can annihilate via a number of channels. For example, they may interact with neutral atoms through charge exchange. In this process, positrons with sufficient energy to overcome the difference between the binding energy of the atom and the binding energy of positronium first strip the valence electrons from atomic nuclei, then form positronium before subsequently annihilating\footnote{Annihilation lifetimes of positronium are $1.2\times10^{-10}\,\mathrm{s}$ for para-positronium, and $1.4\times 10^{-7}\,\mathrm{s}$ for ortho-positronium.}. The total spin angular momentum of the positronium bound state governs the number and energy of emitted gamma rays. Positronium formed with the positron and electron with parallel spins produces the three-photon ortho-positronium continuum \citep{Ore49}. Positronium  formed from positron-electron pairs with anti-parallel spins annihilates to produce two gamma rays at $511\,\mathrm{keV}$. In astrophysical environments, the annihilation of singlet positronium or para-positronium results in a Gaussian-shaped emission line centered at $511\,\mathrm{keV}$, whose width is governed either by the temperature of the medium for thermalized positrons, the residual kinetic energy of the positron for positrons annihilating in flight, or by large scale gas dynamics (kinetic broadening).\\
To form positronium via charge exchange, the positron energy must exceed the difference between the binding energy of positronium ($6.8\,\mathrm{eV}$) and the first ionization potential of the atom (e.g. $w_{IP} = 13.6\,\mathrm{eV}$ for hydrogen), that is, $w\geq w_{IP}-6.8\,\mathrm{eV}$, where $w$ is the positron energy. In the case of charge exchange with hydrogen (helium), the mimimum kinetic energy a positron must have to form positronium is $6.8\,\mathrm{eV}$ ($17.7\,\mathrm{eV}$). If the threshold for positronium formation with an atom is negative, positrons can form positronium at any incident energy in charge exchange interactions with that atom. This is the case for the alkali metal atoms, due to their extremely low ionization potentials. \\
Positrons may also form positronium via radiative recombination with free electrons in the ISM. Unlike charge exchange, positrons with any energy may undergo radiative recombination. Annihilation via radiative recombination also results in the emission of a superimposed para-positronium Gaussian centered at $511\,\mathrm{keV}$ and an additional three-photon ortho-positronium continuum. The temperature of the medium strongly affects the width and shape of the resulting annihilation spectrum \citep{Guessoum05}, as both the fraction of positrons forming positronium and the shape of the $511\,\mathrm{keV}$ line depend on the ISM temperature. Finally, positrons may undergo direct annihilation either through interactions with free electrons in the ISM, or may directly annihilate with electrons bound to atoms. This process results in emission of a Gaussian spectrum where the characteristic width of the line is determined by the ISM electron temperature.\\
Analysis of the measured gamma ray spectrum suggests that the majority of positrons annihilate via charge exchange interactions with neutral atoms, assumed to be hydrogen based on ISM composition \citep{Churazov05}.  The characteristic orthopositronium continuum emission and narrow emission line centered at $511\,\mathrm{keV}$ observed by \textit{INTEGRAL}/SPI indicates that positrons annihilate in a mostly neutral, $\sim10^4\,\mathrm{K}$ ISM \citep{Churazov05,Churazov11, Siegert16}. Simulations and experiment suggest that $90 - 98\%$ of positrons will annihilate via charge exchange during the in-flight phase in these ISM conditions \citep{Guessoum05,BRD79,Wallyn94}. The remaining positrons will thermalize with the surrounding ISM \citep{Prantzos11} before annihilating. It is usually assumed that these thermalized positrons annihilate instantaneously, or on a timescale shorter than the energy loss timescale if the production of positrons is in steady state with respect to the annihilation rate. However, if the annihilation timescale for the thermalized positrons exceeds the slowing down timescale, it is possible for a `reservoir' of low energy positrons to form. In this scenario, it is possible for the current annihlation rate to exceed the positron injection rate. This opens the door on a scenario where positrons are injected into the Galaxy in some kind of outburst event such as those described in \cite{Totani06,Cheng97}.\\
		In this paper we consider how alkali metals in the ISM substantially reduces the annihilation timescale for thermalized positrons in the ISM, and the potential importance of including alkali metal atoms in simulations of positron transport at low energies in the ISM. We also show how including the alkali metal atoms as annihilation targets closes a door on positron production scenarios which do not occur in a steady state.
		
	\section{Positron annihilation and alkali metal atoms}
\begin{figure}
	 		\includegraphics[scale=0.2]{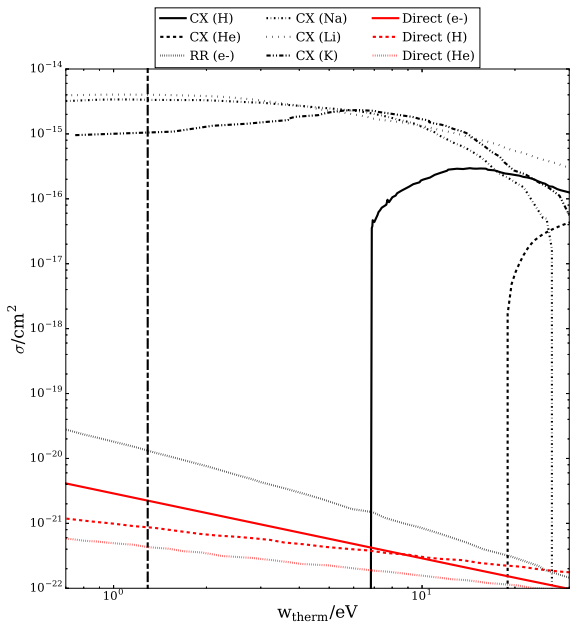}
	 		\caption{\label{fig:xsec} Positronium formation and charge-exchange (CX) cross-section for positron annihilation through various channels as a function of incident positron energy. Hydrogen and helium CX cross-sections based on fits to available data, radiative recombination with free electrons from R. Sutherland (priv. comm.) lithium and sodium CX cross-sections from \citep{kadyrov2016} and potassium CX cross-sections from \cite{Potassiumps}. Direct annihilation cross-sections for positron interactions with free and bound electrons are shown in red. The black dashed line shows the average energy of positrons that have thermalized in the WPIM.}
	 	\end{figure}

\begin{figure}
	 			\includegraphics[scale=0.25]{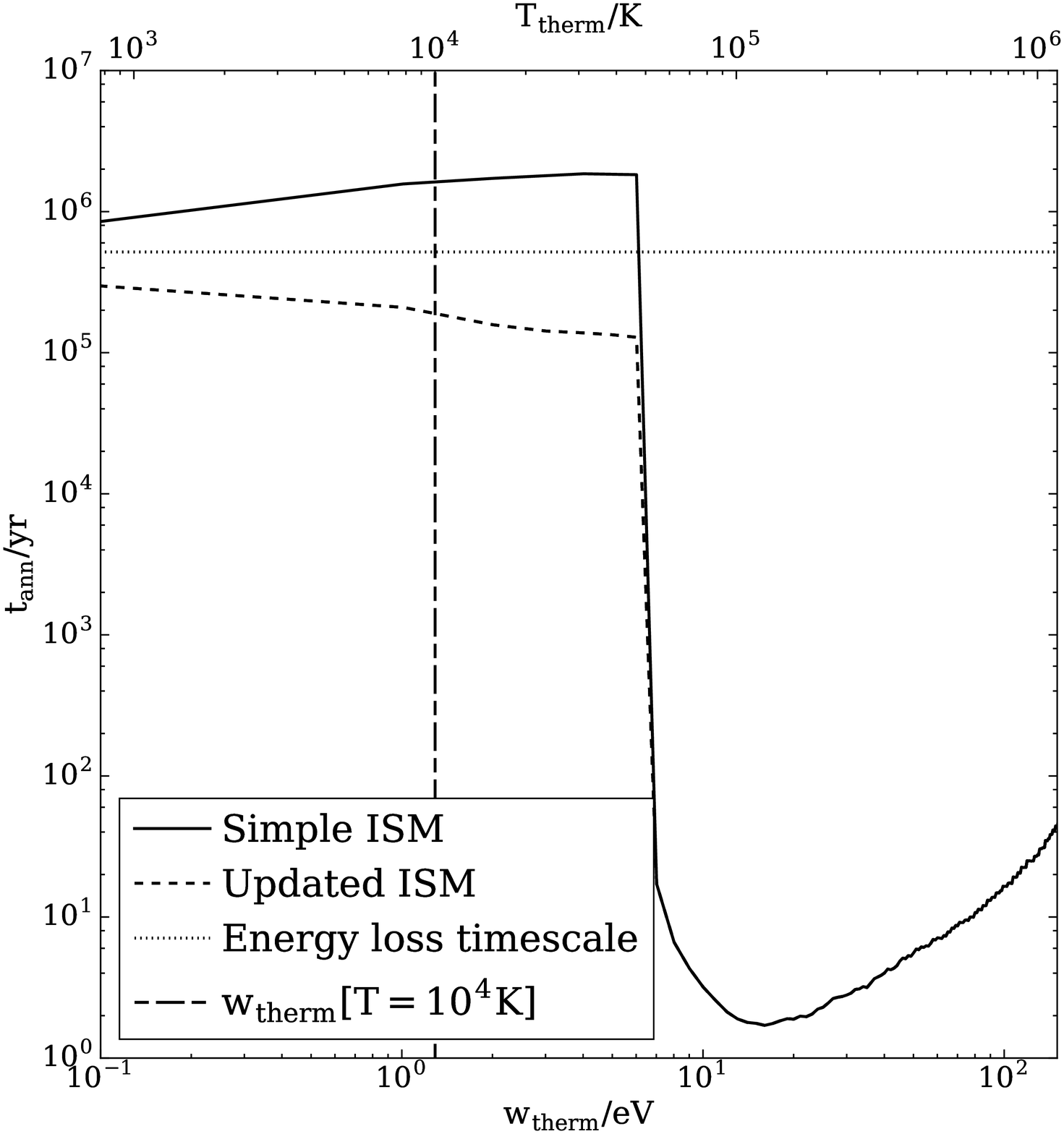}
	 			\caption{\label{fig:timescales} Annihilation timescale for thermalized positrons as a function of positron energy. Including the effect of positron interactions with alkali metal atoms, even at solar metallicity, substantially decreases the annihilation timescale for positrons that thermalize with energies $<6.8\,\mathrm{eV}$. The black dashed line shows the average energy of positrons that have thermalized in the WPIM.}
	 		\end{figure}

	In existing simulations and theory of positron annihilation in the ISM e.g. \citep{Prantzos06,Higdon09,Jean09,Churazov11,Martin2012,Alexis14,Panther17}, positron interactions with hydrogen (both HI and H$_2$), helium, and free electrons are considered (henceforth the `simplified ISM model'). Also of potential importance are dust grains and polycylic aromatic hydrocarbon (PAH) molecules, although the cross-sections and mechanisms associated with annihilation on these species are uncertain \citep{Guessoumdust}, and we do not consider them further in this work.\\
    Hydrogen and helium make up the majority of the diffuse ISM by number density at solar metallicity \citep{Lodders03}, while the free electron density is strongly dependent on the temperature of the medium. Herein, we consider the ISM at $T\sim10^4\,\mathrm{K}$, the ISM temperature at which most positrons seem to annihilate. In this warm, partially ionized phase (WPIM) where the majority of positrons are thought to annihilate \cite{Churazov05,Jean09,Churazov11,Siegert16}, the number density of molecular hydrogen (found in the cold neutral phase) is negligible so we only consider atomic species. The composition and ionization state of the ISM is based on that for the warm ISM described in \citep{Guessoum05}. A summary of the number densities of the considered species in the warm, partially ionized phase of the ISM is given in Table \ref{tab:table1}.\\
    \begin{table}
					\caption{\label{tab:table1}Density of ISM species (per $\mathrm{cm^{3}}$) in the warm, partially ionized phase of the ISM based on \citep{Guessoum05}.}
					\begin{ruledtabular}
						\begin{tabular}{cc}
							Species & Density in WPIM / cm$^{-3}$ \\
							\hline
							Neutral hydrogen & $2.68\times10^{-1}$ \\
							Neutral helium & $2.60\times10^{-2}$ \\
							Free electrons & $2.71\times10^{-3}$\\
							Neutral lithium & $6.05\times10^{-10}$ \\
							Neutral sodium & $6.35\times10^{-7}$ \\
                            Neutral potassium & $4.26\times10^{-8}$ \\
						\end{tabular}
					\end{ruledtabular}
				\end{table}
     In this work, we utilize positronium cross-sections with approximate shape and magnitude determined from fits to the available data \citep{campeanu_partitioning_1987,bailey_internal_2015,kauppila_measurements_1981,machacek_positron_2013,moxom_single_1996,weber_results_1994,stein_measurements_1978,zhou_measurements_1997,laricchia_ionizing_2008,humberston_theoretical_1979,sullivan_excitation_2001,walters_positron_1988,kernoghan_positron_1996,jones_positron_1993,murtagh_positron-impact_2005,charlton_total_1983,charlton_positron_2000} for positron interactions with hydrogen and helium. We find the timescale for positron annihilation at low energies is not affected by use of one dataset preferentially over another for two reasons: firstly, at $10^4\,\mathrm{K}$ the thermalized positron has too little kinetic energy to form positronium in collisions with hydrogen and helium. Consequently, assuming a simplified ISM at these energies positrons only annihilate by direct annihilation with bound electrons in hydrogen and helium atoms, and these cross-sections are comparable to that for direct annihilation with free electrons (see fig \ref{fig:xsec}). Secondly, recent advances in measuring total positron scattering cross-sections have resulted in more precise determination of ionization and excitation cross-sections whereas varying the positronium formation cross-section by $\sim 20\%$ (in accordance with the uncertainties on the combined data set) has a negligible effect on our results\footnote{These advances in measuring total scattering cross-sections will be of importance to positron astrophysics in the context of detailed simulation of collisional transport of positrons in the ISM, where ionization and excitation of atoms in the ISM is responsible for the slowing down of energetic positrons}.\\
	 Due to the non-negative positronium formation thresholds of hydrogen and helium atoms, the annihilation cross-section for positrons interacting with hydrogen (helium) drops to zero at $6.8\,\mathrm{eV}$ ($17.7\,\mathrm{eV}$). In the simplified ISM model, positrons with energies below $6.8\,\mathrm{eV}$ can only annihilate via interactions with free electrons. The cross-section for these interactions is several orders of magnitude lower than that for positron interactions with hydrogen and helium (fig \ref{fig:xsec}). The impact of this decrease in annihilation cross section is interesting in the context of positron annihilation in the ISM, as positrons thermalizing in the ISM phase where most positrons are expected to annihilate (the warm phase) will have energies $<6.8\,\mathrm{eV}$. The lifetime of a thermalized positron (i.e. the time between the positron reaching an energy of $w = 3kT/2$ and subsequently annihilating) with energy $w$ is (e.g. ref \citep{Jean09}) 
	 \begin{equation}
	 \tau_\mathrm{ann} (w) = \bigg(c\beta(w)\sum_{\mathrm{T}}\sigma_\mathrm{T}(w)n_\mathrm{T}\bigg)^{-1}
	 \end{equation}
	 where $c$ is the speed of light, $c\beta$ is the positron velocity, $\sigma_\mathrm{T}$ is the cross section of target species T and $n_\mathrm{T}$ the number density. The impact of the dramatic decrease in the annihilation cross section in the simplified ISM model on the lifetimes of thermalized positrons can be seen as the solid curve in Figure \ref{fig:timescales}.\\
	 In the simplified ISM model, positrons with energies $<6.8\,\mathrm{eV}$ are thought to only form positronium via radiative recombination with free electrons. However, if there are atoms present in the ISM which possess positronium formation thresholds below $6.8\,\mathrm{eV}$, positrons may still form positronium via charge exchange at low energies. Moreover, if the cross-sections of these atoms is sufficiently large, it may also substantially decrease the annihilation timescale for thermalized positrons in the warm ISM. The formation of positronium from positron-alkali scattering is known, and the cross sections are known to an order of magnitude. Of particular interest are the alkali metal sodium\citep{kadyrov2016}, which has large cross sections (on the order of $\sim10^{-15}\,\mathrm{cm^{-2}}$) for charge exchange with positrons at energies $<6.8\,\mathrm{eV}$. For completeness we consider lithium and potassium - however, their inclusion has a negligible effect as their abundances orders of magnitude lower than that of sodium using the abundance tables of \cite{Lodders03}. In the case of potassium, we find its inclusion does not affect the annihilation timescale significantly as its charge exchange cross-section is around half an order of magnitude lower than that of sodium and lithium at the energies considered \citep{Potassiumps}, and its solar abundance around an order of magnitude lower than that of sodium \citep{Lodders03}.\\
	The large cross-section of sodium atoms has a significant impact on the lifetime of positrons with energies $<6.8\,\mathrm{eV}$ despite the comparably low abundance of sodium compared to that of hydrogen and helium (assuming solar abundance in the diffuse ISM). A summary of the species densities in the ISM phase where most positrons appear to annihilate is given in Table \ref{tab:table1}.\\
		In this ISM phase, positrons will thermalize with an energy of 	$w_\mathrm{therm} = 3kT/2 \sim 1.3\,\mathrm{eV}(T/10^4\,\mathrm{K})$. At $w = 1.3\,\mathrm{eV}$, the lifetime of a thermalized positron can be written in parameterized form as
        \begin{align}
\tau = 1.7\times 10^5 \, \frac{0.27 \,\mathrm{cm^{-3}}}{n_H}\bigg(\frac{1-X_H}{0.99}0.085+\frac{\zeta_\mathrm{He}}{0.096}0.004\\+
\frac{\zeta_{Li}}{2.23\times10^{-9}}8.9\times10^{-4}+\frac{\zeta_\mathrm{Na}}{2.34\times10^{-6}}0.78\\+\frac{\zeta_\mathrm{K}}{1.57\times10^{-7}}2.4\times10^{-4}+\frac{X_\mathrm{H}}{3\times10^{-3}}0.0015\bigg)^{-1}\,\mathrm{yr}
        \end{align}
where $n_\mathrm{H}$ is the total hydrogen density, $X_\mathrm{H}$ is the ionization fraction for hydrogen, and the factors $\zeta_\mathrm{T}$ are the abundances of target atom T relative to protosolar abundance \citep{Lodders03}. At $1.3\,\mathrm{eV}$, the lifetime of positrons in the simplified ISM model is $\tau_{1.3} =1.67\,\mathrm{Myr}$, figure \ref{fig:timescales}. Including sodium at solar abundance, along with all species listed in Table \ref{tab:table1} ($\zeta_\mathrm{Na}=2.24\times 10^{-6}$ relative to hydrogen \citep{Lodders03}) reduces the lifetime of such thermalized positrons to $\tau_{1.3}=0.19\,\mathrm{Myr}$.\\
 In particular, in the simplified ISM model, the annihilation timescale exceeds the energy loss timescale for MeV positrons. This apparently allows for the possibility of a low-energy positron `reservoir' that could act to maintain the total number of ISM positrons in the Galaxy in a steady state even if the positrons are injected by large-scale events (e.g. outbursts associated with the supermassive black hole (SMBH) at the Galactic Center [24, 25]) with a typical periodicity significantly in excess of 1 Myr. However, accounting for alkali metal atoms as annihilation targets in a realistic ISM model, closes the door on this possibility: the positrons we detect currently annihilating in the ISM must have been created within the last $\sim\mathrm{Myr}$\footnote{Even this timescale is probably excluded by combination of the observed distribution of positrons over $\gg \mathrm{kpc}$ size scales throughout both the bulge and disk and the fact that large-scale ($>\mathrm{kpc}$) transport of positrons has been ruled out by numerical studies of positron transport \cite{Jean09,Martin2012,Alexis14,Panther17}, even where large scale gas dynamics is invoked \cite{Panther17}. In other words, to maintain the inferred smooth and large-scale distribution of positrons across the Galaxy, each $\sim \mathrm{kpc}$-radius patch must have experienced a positron injection event within the last $< \mathrm{Myr}$ suggesting that the mean periodicity of positron injection events across the Milky Way is $\gg \mathrm{Myr}$.}.\\
	\section{Conclusions}
	In simple ISM models developed to understand the annihilation of the some $5\times10^{43}$ positrons per second in the diffuse ISM of the Milky Way, it is assumed that positrons with kinetic energies below the charge exchange threshold with neutral hydrogen ($<6.8\,\mathrm{eV}$) can only undergo annihilation through interactions with free electrons, or via direct annihilation with electrons bound to hydrogen and helium atoms \citep{Guessoum05,Jean09,Churazov11,Martin2012,Alexis14,Panther17}. The consequence of this assumption is that the calculated lifetime for positrons that thermalize in a $10^4\,\mathrm{K}$ ISM, thought to be between 2 and 10\% of positrons that annihilate in the diffuse ISM, is in excess of $1\,\mathrm{Myr}$. Despite its low abundance at solar metallicity, the inclusion of positron interactions with  sodium can reduce the lifetime of thermalized positrons in the ISM to $\sim 0.1\,\mathrm{Myr}$, and moreover allows a new channel for positrons to form positronium at energies where positronium formation was only assumed to occur through radiative recombination with free electrons. Finally, we conclude that the number of positrons in the Galactic ISM could be maintained in steady-state by injection events with typical periodicity longer than $> \mathrm{Myr}$. 
	\section{Acknowledgements}
    FHP is supported by an Australian Government Research Training Program (RTP) Scholarship. FHP thanks Ralph Sutherland for the supply of updated annihilation cross-sections for radiative recombination. IRS was supported by the Australian Research Council Grant FT160100028. JRM acknowledges the support of the Australian Research Council’s Discovery Early Career Research Award (DECRA) Fellowship.
\appendix*
\section{Appendix:  Annihilation in the multi-phase ISM}
    \begin{table*}
					\caption{\label{tab:table2}Density of ISM species (per $\mathrm{cm^{3}}$) in the warm phase of the ISM based on \citep{Guessoum05}.}
					\begin{ruledtabular}
						\begin{tabular}{ccccc}
							Species & Density in HIM / cm$^{-3}$ & Density in WIM / cm$^{-3}$ & Density in WNM / cm$^{-3}$ & $w_\mathrm{IP}$/eV\\
							\hline
							H$_0$ & 0 & 0 & 0.34 & 13.6\\
							He$_0$ & 0 & $5.60\times10^{-3}$&$3.20\times10^{-2}$& 24.5 \\
                            He$_+$& $7.66\times10^{-5}$ &$5.60\times10^{-3}$&0&54.4\\
							e$^-$ & $1.75\times10^{-3}$ & 0.12&0& 0\\
							Li$_0$ & $3.56\times10^{-12}$ & $2.60\times10^{-10}$& $7.52\times10^{-10}$ & 5.39\\
							Na$_0$ & $3.73\times10^{-9}$ &$2.73\times10^{-7}$ &$7.89\times10^{-7}$& 5.13\\
                            K$_0$ & $2.50\times10^{-12}$& $1.83\times10^{-8}$&$5.30\times10^{-8}$&4.3407\\
                            \hline
                         	H$_+$ & $1.59\times10^{-3}$ & 0.11& 0 &-\\
                            He$_2+$ & $7.66\times10^{-5}$& 0 &0 &-\\
						\end{tabular}
					\end{ruledtabular}
				\end{table*}
	 		\begin{figure*}
	 			\includegraphics[width = \textwidth]{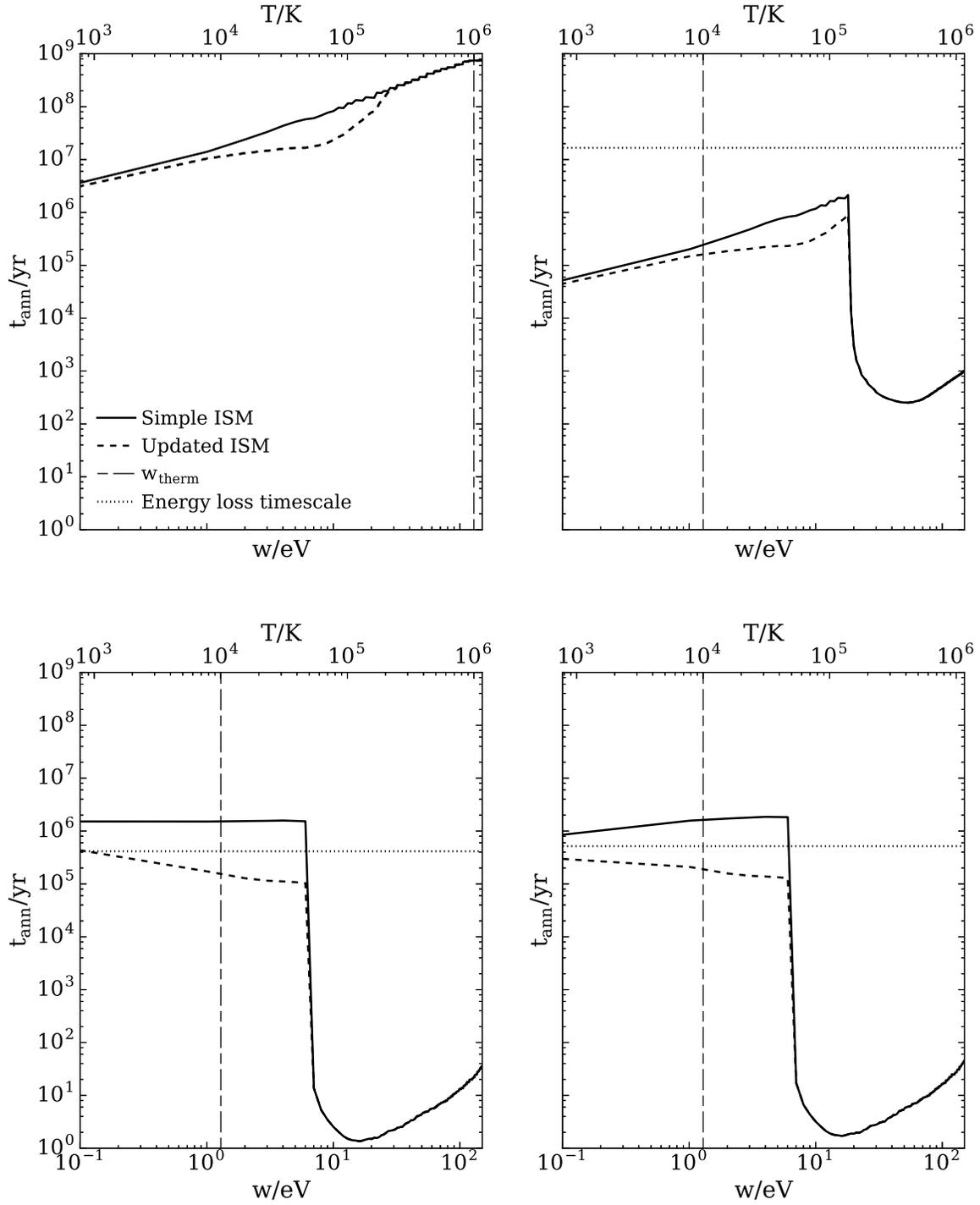}
	 			\caption{\label{fig:timescalesmultiphase} Comparison of annihilation timescales for thermalized positrons in the HIM (top left), WIM (top right), WNM (bottom left) and WPIM (bottom right). Vertical dashed lines indicate the average energy of positrons thermalizing in each phase. Horizontal lines show the energy loss timescale for MeV positrons in each phase. In the case of the HIM (top left), the energy loss timescale for MeV positrons is $\sim 10^{11}\,\mathrm{yr}$}
	 		\end{figure*}
In Fig \ref{fig:timescalesmultiphase}, we show that the inclusion of alkali metal atoms, most notably sodium, reduces the annihilation timescale for thermalized positrons in all phases of the ISM. However, this is most significant in the case of the WNM and WPIM where the majority of positrons are thought to annihilate via interactions with neutral atoms \cite{Churazov05}. In these phases, where $T\sim 10^{4}\,\mathrm{K}$, positrons thermalize at $\sim 1.3\,\mathrm{eV}$. At this energy, the annihilation timescale for positrons in the simple ISM is significantly longer than the slowing down time for the positrons. Including alkali metal atoms as annihilation targets for thermalized positrons reduces the annihilation timescale, and thus for the updated ISM model we find $\tau_\mathrm{ann}<\tau_\mathrm{loss}$, with the consequences being the same as those discussed in the main article.\\
While the observed positron annihilation spectrum suggests that positrons annihilate in a warm, partially ionized ISM, the possibility that positrons annihilate in multiple phases of the ISM is not excluded \cite{Churazov05}. However, the characteristics of the annihilation put strong constraints on the proportion of positrons annihilating in each phase - hot ionised medium (HIM), warm ionized medium (WIM), warm neutral medium (WNM) and cold molecular medium (CMM). In \cite{Churazov05} it is found that annihilation in a multiphase medium can explain the observed annihilation spectrum only where no more than ∼ 8\% of positrons annihilate in the hot ionized phase ($T \geq 10^6\,\mathrm{K}$). Furthermore annihilation in CMM ($T\leq10^3\,\mathrm{K}$) cannot make a dominant contribution to the annihilation spectrum in the presence of a multiphase medium. Moreover, the presence of the positronium continuum suggests that $>95\%$ of positrons annihilate via interactions with neutral atoms (usually assumed to be hydrogen and helium). This would suggest that the majority of positrons likely annihilate in the WNM. For completeness, the densities of the species in our updated ISM model are shown in Table \ref{tab:table2} for each of the HIM, WIM and WNM. We do not consider the CMM based on the constraints on the annihilation spectrum - annihilation on molecular hydrogen in the CMM results in a broad emission line which is inconsistent with the spectrum observed by SPI \cite{Churazov05,Churazov11,Siegert16} - and moreover the spatial morphology of the positron annihilation signal, which is significantly more extended than the distribution of cold molecular gas in the Milky Way.\\
\bibliography{PantherBib}

\end{document}